# Frequency modulation Fourier transform spectroscopy

Julien Mandon, Guy Guelachvili, Nathalie Picqué

Laboratoire de Photophysique Moléculaire, CNRS; Univ. Paris-Sud, Bâtiment 350, 91405

Orsay, France

Corresponding author:
Dr. Nathalie Picqué,
Laboratoire de Photophysique Moléculaire
Unité Propre du CNRS, Université Paris Sud, Bâtiment 350
91405 Orsay Cedex, France
Phone number: +33 1 69156649
Fax number: +33 1 69157530
Email: nathalie.picque@ppm.u-psud.fr
Web: http://www.laser-fts.org

Abstract: A new method, FM-FTS, combining Frequency Modulation heterodyne laser spectroscopy and Fourier Transform Spectroscopy is presented. It provides simultaneous sensitive measurement of absorption and dispersion profiles with broadband spectral coverage capabilities. Experimental demonstration is made on the overtone spectrum of $C_2H_2$ in the 1.5 µm region.

120.6200 Spectrometers and spectroscopic instrumentation, 300.6300 Spectroscopy, Fourier transforms, 300.6380 Spectroscopy, modulation, 300.6360 Spectroscopy, laser, 300.6310 Spectroscopy, heterodyne, 300.6390 Spectroscopy, molecular, 120.5060 Phase modulation

Improving sensitivity is presently one of the major concern of spectroscopists. This may be obtained both from the enhancement of the intrinsic signal, and from the reduction of the background noise. In this latter case, modulation has been one of the most effective approach. In particular, Frequency Modulation (FM) absorption spectroscopy [1] has reached detection sensitivity near to the fundamental quantum noise limit, by shifting the frequency modulation of the measurements to a frequency range where the 1/f noise becomes negligible. Moreover, FM spectroscopy benefits from high-speed detection and simultaneous measurement of absorption and dispersion signals. Since Bjorklund's first demonstrations [1,2] of the efficiency of FM spectroscopy with a single-mode continuous-wave dye laser, the technique has been widely used as a tunable laser spectroscopic method in fields such as laser stabilization [3], two-photon spectroscopy [4], optical heterodyne saturation spectroscopy [5], trace gas detection [6]. In most schemes, the laser wavelength is scanned across the atomic/molecular resonance to retrieve the line shape. More rarely, the modulation frequency is tuned. However in both cases, the measurements are limited to narrow spectral ranges.

This letter reports the first results in FM broadband spectroscopy. This work is motivated by our ongoing effort of implementing a new spectroscopic approach simultaneously delivering sensitivity, resolution, accuracy, broad spectral coverage and rapid acquisition. The basic idea, named FM-FTS, is to associate the advantages of FM spectroscopy and high-resolution Fourier transform spectroscopy (FTS). FTS is able to record at once extended ranges, with no spectral restriction. In particular it gives easy access to the infrared domain. In this letter, a new way of modulating the interferogram is implemented. The key concept is that a radio frequency (RF) modulation is performed. The beat signal at the output port of the Fourier transform spectrometer is modulated at constant RF, which is about $10^4$ times greater than the audio frequency generally delivered by the interferometer optical conversion. Together with the advantage, over classical FTS, of measurements performed at much higher frequency, our approach benefits from the synchronous detection ability and from the simultaneous acquisition of both the absorption and the dispersion of the recorded profiles.

The experimental principle is presented in Fig. 1. The light emitted by the broadband source is first passing through the interferometer. The output beam is then phase-modulated by an electro-optic modulator (EOM) before entering the absorption cell and falling on the fast detector. The synchronous detection of the detector signal is realized by the lock-in amplifier at the EOM driver reference frequency $f_m$. Recorded data are finally stored on the computer disk with their corresponding path difference position $\Delta$. Their Fourier transform is the spectrum.

In more details, the electric field $\mathbf{E}$ at the output of the interferometer may be written as:

$$\mathbf{E}(\Delta, t) = \int \frac{\mathbf{E}_0(\omega_c)}{2}\left[1 + \exp\left(-i\omega_c\frac{\Delta}{c}\right)\right]\exp(i\omega_c t)\mathrm{d}\omega_c + \text{c.c.} \quad (1)$$

where $\mathbf{E_0}$ is the electric field amplitude of the source at $\omega_c$ optical pulsation, $c$ is the velocity of light and c.c the conjugate complex of the preceding expression in Eq. 1. The EOM effect on the beam is assumed to have a low modulation index M. As a consequence, each carrier wave of pulsation $\omega_c$, has two weak sidebands located at $\pm\,\omega_m = \pm\,2\pi\,f_m$. Equation (1) becomes:

$$\mathbf{E}(\Delta, t) = \int \frac{\mathbf{E}_0(\omega_c)}{2}\left[1 + \exp\left(-i\omega_c\frac{\Delta}{c}\right)\right]\left\{\exp\left(i\omega_c t\right)\right.$$

$$\left. + \text{M}\exp\left[i\left(\omega_c + \omega_m\right)t\right] - \text{M}\exp\left[i\left(\omega_c - \omega_m\right)t\right]\right\}\mathrm{d}\omega_c + \text{c.c.} \quad (2)$$

When interacting with the gas, the carrier and the sidebands experience attenuation and phase-shift due to absorption and dispersion. Following the notations introduced in [1], this interaction may be written as $\exp(-\delta(\omega)- i\ \phi(\omega))$ where $\delta$ is the amplitude attenuation and $\phi$ is the phase shift. The following convention is adopted: $\delta_n$ and $\phi_n$ denotes for n = 0, ±1 the respective components at $\omega_c$ and $\omega_c \pm \omega_m$. Then Eq. 2 may be written:

$$\mathbf{E}(\Delta, t) = \int \frac{\mathbf{E}_0(\omega_c)}{2} \left[ 1 + \exp\left(-i\omega_c \frac{\Delta}{c}\right) \right] \left\{ \exp\left(-\delta_0 - i\phi_0\right) \exp\left(i\omega_c t\right) \right.$$

$$+ M \exp\left(-\delta_{+1} - i\phi_{+1}\right) \exp\left[i\left(\omega_c + \omega_m\right)t\right] - M \exp\left(-\delta_{-1} - i\phi_{-1}\right) \exp\left[i\left(\omega_c - \omega_m\right)t\right] \right\} d\omega_c + \text{c.c.} \quad (3)$$

The intensity I detected by the fast photodetector is proportional to :

$$\mathrm{I}(\Delta, t) \propto \mathbf{E}(\Delta, t)\mathbf{E}^*(\Delta, t). \quad (4)$$

$$\mathrm{I}(\Delta, t) \propto \int \left( \left[ \exp\left(-2\delta_0\right) + \exp\left(-2\delta_{+1}\right) + \exp\left(-2\delta_{-1}\right) \right] \left[ 1 + \cos\left(\omega_c \frac{\Delta}{c}\right) \right] \right.$$

$$+ 2M \cos\left(\omega_m t\right) \left[ \exp\left(-\delta_0 - \delta_{+1}\right) \cos\left(\phi_0 - \phi_{+1}\right) - \exp\left(-\delta_0 - \delta_{-1}\right) \cos\left(\phi_0 - \phi_{-1}\right) \right] \left[ 1 + \cos\left(\omega_c \frac{\Delta}{c}\right) \right]$$

$$+ 2M \sin\left(\omega_m t\right) \left[ \exp\left(-\delta_0 - \delta_{-1}\right) \sin\left(\phi_{-1} - \phi_0\right) - \exp\left(-\delta_0 - \delta_{+1}\right) \sin\left(\phi_0 - \phi_{+1}\right) \right] \left[ 1 + \cos\left(\omega_c \frac{\Delta}{c}\right) \right]$$

$$+ 2M^2 \cos\left(2\omega_m t\right) \left[ \exp\left(-\delta_{+1} - \delta_{-1}\right) \sin\left(\phi_{-1} - \phi_{+1}\right) \right] \left[ 1 + \cos\left(\omega_c \frac{\Delta}{c}\right) \right]$$

$$+ 2M^2 \sin\left(2\omega_m t\right) \left[ \exp\left(-\delta_{+1} - \delta_{-1}\right) \cos\left(\phi_{-1} - \phi_{+1}\right) \right] \left[ 1 + \cos\left(\omega_c \frac{\Delta}{c}\right) \right] \right) d\omega_c. \quad (5)$$

After synchronous detection at $f_m$ frequency and with the assumption that $|\delta_0 - \delta_j| \ll 1$ and $|\phi_0 - \phi_j| \ll 1$ (with $j = \pm 1$), the in-phase $\mathrm{I}_{\cos}(\Delta)$ and the in-quadrature $\mathrm{I}_{\sin}(\Delta)$ parts of the electric signal are given by

$$\mathrm{I}_{\cos}(\Delta) \propto M \int \left[ 1 + \cos\left(\omega_c \frac{\Delta}{c}\right) \right] \exp\left(-2\delta_0\right)\left(\delta_{-1} - \delta_{+1}\right) d\omega_c. \quad (6)$$

$$\mathrm{I}_{\sin}(\Delta) \propto M \int \left[ 1 + \cos\left(\omega_c \frac{\Delta}{c}\right) \right] \exp\left(-2\delta_0\right)\left(\phi_{+1} + \phi_{-1} - 2\phi_0\right) d\omega_c. \quad (7)$$

Summarizing, two interferograms are simultaneously measured, allowing to obtain broadband FM spectra. The in-phase interferogram provides spectrally resolved information on the difference of absorption experienced by each group of two sidebands. The in-quadrature interferogram gives the difference between the average of the dispersions experienced by the sidebands and the dispersion undergone by each carrier.

For this first experimental demonstration, a narrow-band emission source covering 0.25 cm$^{-1}$ (7.5 GHz) has been implemented as a test source. It is made of a fiber-coupled distributed feedback laser diode emitting around 1530 nm with an output power of a few mW. The current of the laser diode is modulated at about 20 Hz by a ramp generator. At each path difference step, while the interferometer is recording one interferogram sample, the laser frequency excursion is equal to 7.5 GHz, corresponding to one period of the triangular ramp. Consequently, for the interferometer, the laser diode behaves as a continuous emission source emitting over 0.25 cm$^{-1}$. The interferometer output light is phase-modulated at $f_m = 150$ MHz by the EOM and passes through an 80-cm cell filled at 10 hPa with acetylene in natural abundance. The light is next focused on an InGaAs nanosecond infrared photodetector, which according to Eq.5 delivers a signal proportional to the intensity of the beam containing a beat signal at the RF modulation frequency. The amplified detector signal is mixed with the reference signal at $f_m$, down to d.c., using a commercial high frequency dual-phase lock-in amplifier. The reference may be phase-shifted with respect the signal used to drive the EOM. The two channels detected in-phase and in-quadrature are measured simultaneously.

Figure 2 shows a typical in-phase interferogram of $C_2H_2$. Its shape is characteristic of an interferogram of first-derivative type line-shapes. The 3 cm period amplitude modulation is due to the beat between the two strongest acetylene lines in the explored spectral domain. Figure 3 shows the two narrow-band spectra, Fourier transform of the in-phase (absorption) and in-quadrature (dispersion) interferograms. The spectral domain extension is limited by the tuning capabilities of the diode laser, which was used as a test source. This does not restrict the generality of the present demonstration. The lines belong to the $\nu_1+\nu_3$ and $\nu_1+\nu_3+\nu_5^1-\nu_5^1$ overtone bands of $^{12}C_2H_2$. The unapodised spectral instrumental resolution: $12.5 \ 10^{-3} \ cm^{-1}$ (0.375 GHz) is narrower than the Doppler width of the lines. Signal to noise ratio is of the order of 1200. The total recording time of the order of 15 minutes is due to the need of adapting the interferometer recording mode procedure to the rather low laser diode frequency excursion period.

The present validation of FM-FTS with a narrow band light source made the experience much simpler. Indeed, in wideband FTS, processing the signal of the interferogram needs special dynamic range solutions. Thanks to the only $0.25 \ cm^{-1}$-wide spectrum analysed in this experiment, a sophisticate RF detection chain, presently under development, was not necessary. The design of our Connes-type interferometer allows a balanced detection of the signals recorded at the two output ports. This will be helpful to remove the part of the interferogram which is not modulated by path difference and to consequently improve the dynamic range of the measurements. Similar solutions have already been successfully practiced for time-resolved FTS [7]. In FM-FTS, they are formally even easier to implement since the signal may be band-pass filtered around the modulation radio-frequency.

In the present experimental set-up, the light should sequentially reach the equipment parts as shown in Fig.1. Briefly, to have a broadband equivalent of FM tunable laser spectroscopy, the sidebands generated by the EOM must not be resolved by the spectrometer. Also, since each carrier and its sidebands have to experience different attenuation and phase-shift, the EOM must be placed before the cell containing the gas of interest. This matter will be discussed in more detail elsewhere.

This first FM-FTS experiment demonstrates the feasibility of coupling broadband laser sources, Fourier spectrometers and RF detection. This opens new perspectives in high sensitivity multiplex spectroscopy. FM-FTS may be coupled to a large variety of high brightness sources. This includes broadband cw lasers, supercontinua sources, mode-locked lasers as demonstrated recently [8], and Amplified Spontaneous Emission sources. Frequency nonlinear conversion may also be used when no laser source is available in the spectral range of interest.

FM-FTS may be practiced with any kind of Fourier transform spectrometers, including commercially available instruments, at the expense of reasonable modifications in the signal detection scheme. The approach is also suitable at low spectral resolution. In such case, modulation frequencies lying in the GHz domain may be used. Moreover, FM-FTS induces new practices in Fourier transform spectroscopy. The modulation frequency is very high. The optical fringes generated by the interferometer can then be scanned at a much higher frequency than what is usually practiced nowadays. Path difference variation of the order of 1 m/s, is easily affordable. It corresponds to acquisition times expressed in second when presently the most efficient existing high resolution interferometers need 1 to 10 hours to record interferograms. Additionally, due to the low étendue of the analysed laser beams in our method, miniaturized instruments may be implemented. In addition to the radio-frequency detection scheme, sensitivity may be further enhanced by using an external optical resonator, thus increasing the effective absorption length.

With FM-FTS, both the absorption and the dispersion associated with each spectral features are measured simultaneously. Despite its recognized interest for lineshape parameters retrieval, traditional dispersion spectroscopy has been poorly developed, only at low spectral resolution, mostly due to its experimental complexity. FM-FTS should represent an easy manner of getting this information over extended spectral domains, which may induce new interest to the experimental investigation of dispersion profiles.

Figure captions

Fig. 1.
Schematic of the experimental setup.

Fig. 2.
Absorption interferogram using in-phase RF detection with FM-FTS. Maximum path difference is 40 cm corresponding to $12.5 \ 10^{-3} \ cm^{-1}$ unapodized resolution.

Fig. 3.
FM-FTS dispersion and absorption spectra of the acetylene molecule at 1528.6 nm. The middle plot represents the line relative intensities taken from the HITRAN database.

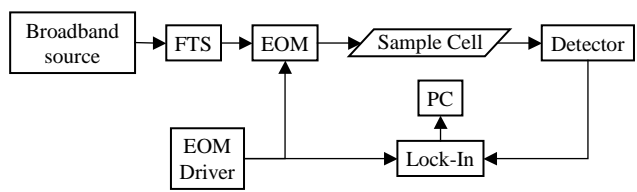

Fig. 1. Schematic of the experimental setup.

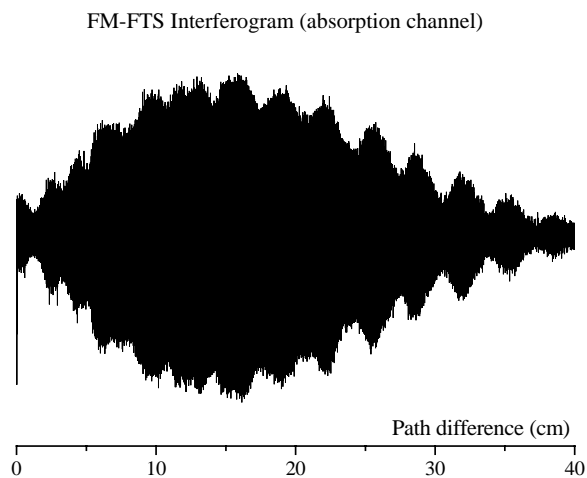

Fig. 2. Absorption interferogram using in-phase RF detection with FM-FTS. Maximum path difference is 40 cm corresponding to 12.5 $10^{-3}$ cm$^{-1}$ unapodized resolution.

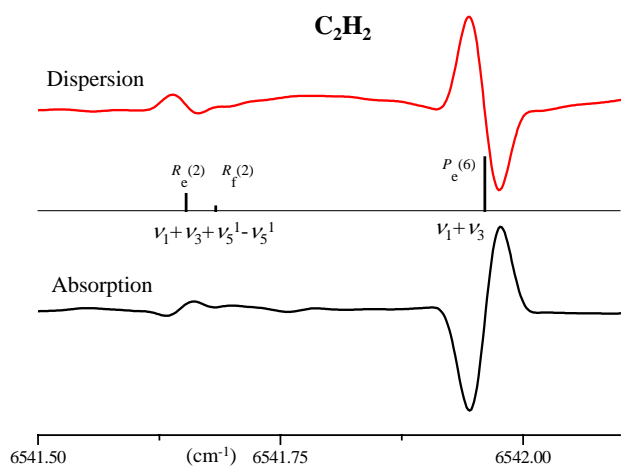

Fig. 3. FM-FTS dispersion and absorption spectra of the acetylene molecule at 1528.6 nm. The middle plot represents the line relative intensities taken from the HITRAN database.